\documentclass[aps,prl,twocolumn,groupedaddress]{revtex4-1}

\usepackage{graphicx}
\usepackage{color}
\usepackage{amsmath}
\usepackage[T1]{fontenc}
\usepackage[english]{babel}
\usepackage{hyperref}

\renewcommand{\frac}{\dfrac}

\newcommand{\Dm}[1]{\frac{D #1}{D t}}
\newcommand{\grad}{\boldsymbol{\nabla}}
\newcommand{\diver}[1]{\grad \cdot \vec{#1}}
\newcommand{\lap}{\nabla^2}

\renewcommand{\vec}{\mathbf}

\begin{document}


\title{Shock formation in the collapse of a vapor nano-bubble}


\author{F. Magaletti}
\author{L. Marino}
\author{C.M. Casciola}
\email[]{carlomassimo.casciola@uniroma1.it}
\affiliation{Dipartimento di Ingegneria Meccanica e Aerospaziale, 
Universit\`a di Roma {\em{La Sapienza}}, 
Via Eudossiana 18, 00184 Roma {\it{Italy}} 
}


\date{\today}

\begin{abstract}
In this Letter a diffuse-interface model featuring  phase change, transition to supercritical conditions, thermal conduction, compressibility effects and shock wave propagation  is exploited to deal with the dynamics of a cavitation bubble. At variance with previous descriptions, the model is uniformly valid for all phases (liquid, vapor and supercritical) and phase transitions involved,  allowing to describe the non-equilibrium processes ongoing during the collapse.
As consequence of this unitary description,  rather unexpectedly for pure vapor bubbles, the numerical experiments  show  that the collapse  is accompanied by the emission of a strong shock wave in the liquid and by the oscillation of the bubble that periodically disappears and reappears, due to transition to super/sub critical conditions. The mechanism of shock wave formation is strongly related to the transition of the vapor to supercritical state, with a progressive steepening of the compression wave to form the shock which is eventually reflected as an outward propagating wave in the liquid.
\end{abstract}

\maketitle


Vapor bubble collapse is a fascinating classical problem \cite{rayleigh1917} involving vapor-liquid phase transition and  extreme pressures and temperatures \cite{science_nuclear_emission, ohl2010aiming}.  Typical experiments concern ultra-fast imaging and the analysis of light and sound emitted after the collapse \cite{pecha2000microimplosions, sankin2005shock, weninger2000observation}, see also the review \cite{brenner2002single}.
Both free cavitation bubbles and nano bubbles at solid walls are increasingly investigated \cite{brenner_lohse2008dynamic, weijs_lohse2013surface, zhang2013deactivation}. 
In ordinary conditions, gas bubble nucleation is associated with the metastability of the mixture of liquid and dissolved gas which, once the free energy barrier between the two states is overcome and the critical nucleus is formed, evolves toward a finite size bubble. The same mechanism is at work in forming pure vapor bubbles when the (ultra pure) liquid is kept in metastable conditions \cite{davitt2010water}, 
i.e. its pressure is below the equilibrium vapor pressure at the given temperature. In these conditions, away from solid walls (see e.g. \cite{giacomello2013geometry} for the role of asperities on solid surfaces as a catalyst of bubble nucleation), local density fluctuations can generate the critical vapor nucleus from which the eventual bubble is formed.

Intermingled  phenomenologies, \cite{brennen2013cavitation, plesset_review}, such as interface dynamics \cite{keller, plesset1971collapse}, thermodynamics of phase change \cite{fujikawa1980}, and dissolved  gas diffusion \cite{akhatov2001collapse}, are a challenge to theoretical modeling of cavitation.
The available descriptions combine two distinct adjoining regions, liquid and vapor phase, respectively, with vapor pressure taken to be the
saturation pressure  \cite{hao1999dynamics} and phase transition accounted for through suitable kinetic equations and latent heat release \cite{akhatov2001collapse}.

Contrary to available models, the diffuse interface approach discussed in the present Letter encompasses all phases (liquid, vapor and supercritical) and phase transitions involved, embedding capillary forces, compressibility effects and shock wave propagation.
The approach enables an unprecedented analysis of collapse, where the bubble interface speed may exceed  the speed of sound. This leads to the formation of a shock wave focused towards the bubble that is successively reflected back in the liquid.  Latent heat of condensation and rapid compression  locally bring the vapor in supercritical conditions. This explains the observed rebounds usually considered a typical feature of incondensable gas bubbles. Indeed, the liquid-vapor interface may disappear and reappear again, according to the local thermodynamic conditions.

{\bf The physical model.}\label{math_model}
We exploit an unsteady diffuse interface description \cite{mcfadden} of the multiphase flow based on the van der Waals gradient approximation of the free energy functional \cite{dell1995radius, jamet_JCP}
%
\begin{equation}\label{eq.Ffunctional}
F[\rho, \theta] = \int_{\Omega} \left( \hat{f}_0\left(\rho, \theta\right) + \frac{\lambda}{2}\vert \grad \rho \vert^2\right) \,dV \,,
\end{equation}
where $\lambda$ is a coefficient controlling surface tension and interface thickness.
For a van der Waals fluid, the free energy per unit volume at temperature $\theta$ and density $\rho$ is 
\begin{equation}\label{eq.free_en}
\hat{f}_0\left(\rho, \theta\right) =  \bar{R} \rho \theta \left[-1 + \log\left(\frac{\rho\,K\,\theta^\delta}{1 - b\rho}\right)\right] - a \rho^2 \,,
\end{equation}
with $\delta = \bar{R}/c_v$, $\bar{R}$ the gas constant, $c_v$ the constant volume specific heat,
$a$ and $b$ the van der Waals coefficients, with $K$ a suitable constant \cite{ruggeri}.
The  evolution is governed by the  conservation equations for mass $\partial_t{\rho} + \grad \cdot \left(\rho \vec{u}\right) = 0$,
momentum $\partial_t \left( \rho \vec{u}\right) + \grad \cdot \left( \rho \vec{u} \otimes \vec{u} \right) = \grad \cdot \boldsymbol{\tau}$,
and total energy, $\partial_t {E} + \grad \cdot \left( \vec{u}E \right)  =  \grad \cdot \left[ \boldsymbol{\tau}\cdot \vec{u} - \lambda \rho \diver{u}\grad\rho + k \grad\theta\right]$, where $k$ is the thermal conductivity. 
The stress tensor, $\boldsymbol{\tau} = - p_0 \boldsymbol{I} + \boldsymbol{\tau}^v  + \boldsymbol{\tau}^c$,
with $p_0 = \rho^2 \partial ({\hat f_0}/\rho)/\partial \rho\vert_{\theta}$ the pressure and $\boldsymbol{\tau}^v$ the classical viscous component, 
features the diffused capillary stress (see {\sl Supplemental Material} (SM) \cite{SM}),
\begin{eqnarray}
	\nonumber
	\boldsymbol{\tau}^c &=&  \lambda\left[\left(\frac{1}{2}\vert\grad\rho\vert^2 + \rho\lap\rho\right)\boldsymbol{I} - \grad\rho\otimes\grad\rho\right]   \, .
\end{eqnarray}
$\boldsymbol{\tau}^c$ accounts for surface tension $\sigma$, defined starting from the deviatoric component of the equilibrium contribution ($\boldsymbol{\tau}^{eq}  =- p_0 \boldsymbol{I} + \boldsymbol{\tau}^c$) to the stress tensor, $\boldsymbol{\tau}^{dev} =
\boldsymbol{\tau}^{eq} - 1/3 \, {\rm tr}\left(\boldsymbol{\tau}^{eq} \right) = -\lambda \grad \rho \otimes \grad \rho$, as the integral along a one-dimensional cut normal to the interface 
\begin{equation*}\label{eq.surf_tens}
	\sigma = -\int_{-\infty}^{+\infty}  \boldsymbol{\nu}  \cdot  \left(\boldsymbol{\tau}^{dev}  \cdot \boldsymbol{\nu} \right)\, \mathrm{d}n =
	\int_{-\infty}^{+\infty} \lambda \left(\frac{\mathrm{d}\rho}{\mathrm{d}n}\right)^2 \mathrm{d}n\, ,
\end{equation*}
with $n$ the normal distance across the interface, and $\boldsymbol{\nu}$ the unit normal to the cut (tangent to the interface).
In fact, the integral is restricted to the region of thickness $\Delta n \propto \sqrt{\lambda}$ where ${\rm d} \rho/{\rm d}  n$ is substantially concentrated, resulting in $\sigma = \int_{\rho_V}^{\rho_L} \sqrt{2 \lambda w(\rho)} d\rho \propto \sqrt{\lambda}$ \cite{SM, cahn1}, where $L$ and $V$ denote liquid and vapor, respectively.


 
The thermodynamic condition $(\rho,\theta)$ identifies completely the bulk state, and Eq.~(\ref{eq.free_en})  provides the latent heat released during the phase change, $\theta \Delta \eta$, with $\eta = -\partial f_0/\partial \theta\vert_\rho$ the entropy per unit mass. As a consequence, the present model intrinsically features  the phase transition whenever the local thermodynamic conditions are appropriate, as shown by the temperature equation
\begin{equation}
\label{eq.temperatura}
\rho c_v \Dm{\theta} = - \rho \theta \frac{\partial \eta}{\partial \rho}\bigg\vert_{\theta} \Dm{\rho} + \boldsymbol{\tau}^v : \boldsymbol{\nabla} \textbf{u} +
\boldsymbol{\nabla} \cdot \left(k  \boldsymbol{\nabla} \theta\right) \, .
\end{equation}
Along the phase transition, the factor $\theta {\partial \eta}/{\partial \rho}\vert_{\theta}$, basically consists of the amount of  heat required to change the state by $d \rho$. Assuming, for the sake of definiteness, an isothermal transformation in subcritical conditions, the integral of this term from vapor to liquid  state is indeed the latent heat $\theta \Delta \eta$. Consistently with physical expectation, the van der Waals model prescribes a latent heat release only in subcritical conditions.
In supercritical conditions, the same term is related to the isochoric thermal pressure coefficient
$(1/p)(\partial \eta/\partial v)\vert_{\theta}   = (1/p)(\partial p/\partial \theta)\vert_{v}$, with $v$ the specific volume.
Clearly, during the actual evolution, the thermodynamic transformation followed by a material element is not isothermal, the actual transformation being dictated by mass, momentum and energy conservation.

The model can in principle be extended to a more sophisticated equation of state (EoS) than the van der Waals', when a direct comparison with experiments is sought for, see, e.g., \cite{moss1994hydrodynamic} where an empirical EoS is used  to include dissociation and ionization effects which 
become relevant in the late stage of the collapse.

{\bf Simulation details}. \label{num_scheme}
The system of conservation laws to be solved constitutes a quite non-standard problem that calls for specialized numerical techniques. We list here a few
issues to provide a flavor of the numerics used in the simulations, see {\sl Supplemental Material, Section B} \cite{SM} for a more complete discussion of the numerical method: i) The extremely thin liquid-vapor interface calls for a high numerical resolution; ii) Liquid, vapor, and supercritical fluid compressibility gives rise to shock waves propagating in a non-uniform, multiphase environment; iii) The system manifests a compound nature, partly controlled by acoustics (hyperbolic behavior) and partly induced by viscosity and capillarity (diffusion and dispersion, respectively); iv) Although the sound speed, $c^2 = \partial p_0/\partial \rho\vert_{\eta}$, is well defined in most of the phase space ($c^2 > 0$), a region exists below the spinodal where $c^2 <0$. This hybrid behavior \cite{affouf1991numerical}  is cause of  failure for  the standard hyperbolic solvers.

{\bf Bubble dynamics and shock evolution.}\label{num_res} The model described above is used to perform numerical simulations of spherically symmetric, pressure-induced bubble collapse.  A vapor nano-bubble with radius $ R_{eq} = 100 \, {\rm nm}$ is initially in equilibrium and the collapse is initiated by an overpressure enforced on the liquid, $\left(p_\infty - p_e\right)/p_e = \Delta p/p_e > 0$, where $p_e$ is the liquid equilibrium pressure.
Cases differing for overpressure and for thermal conductivity, measured by $Pr = {3} {\mu\, \bar{R}}/(8{k})$, are considered  to analyze their effects on the collapse dynamics. In all cases the initial temperature is $\theta_e/\theta_c = 0.5$  and the  surface tension is $\sigma/(p_c R_{eq})=0.045$, corresponding  to a realistic value  $\sigma = 0.09\, N/m$  for water. Here and throughout the subscript $c$ denotes critical values.
 
Along the evolution, the bubble radius is measured as the distance of the liquid region from the center of the bubble,
 \begin{figure}[t!]
 {\includegraphics[width=0.5\textwidth]{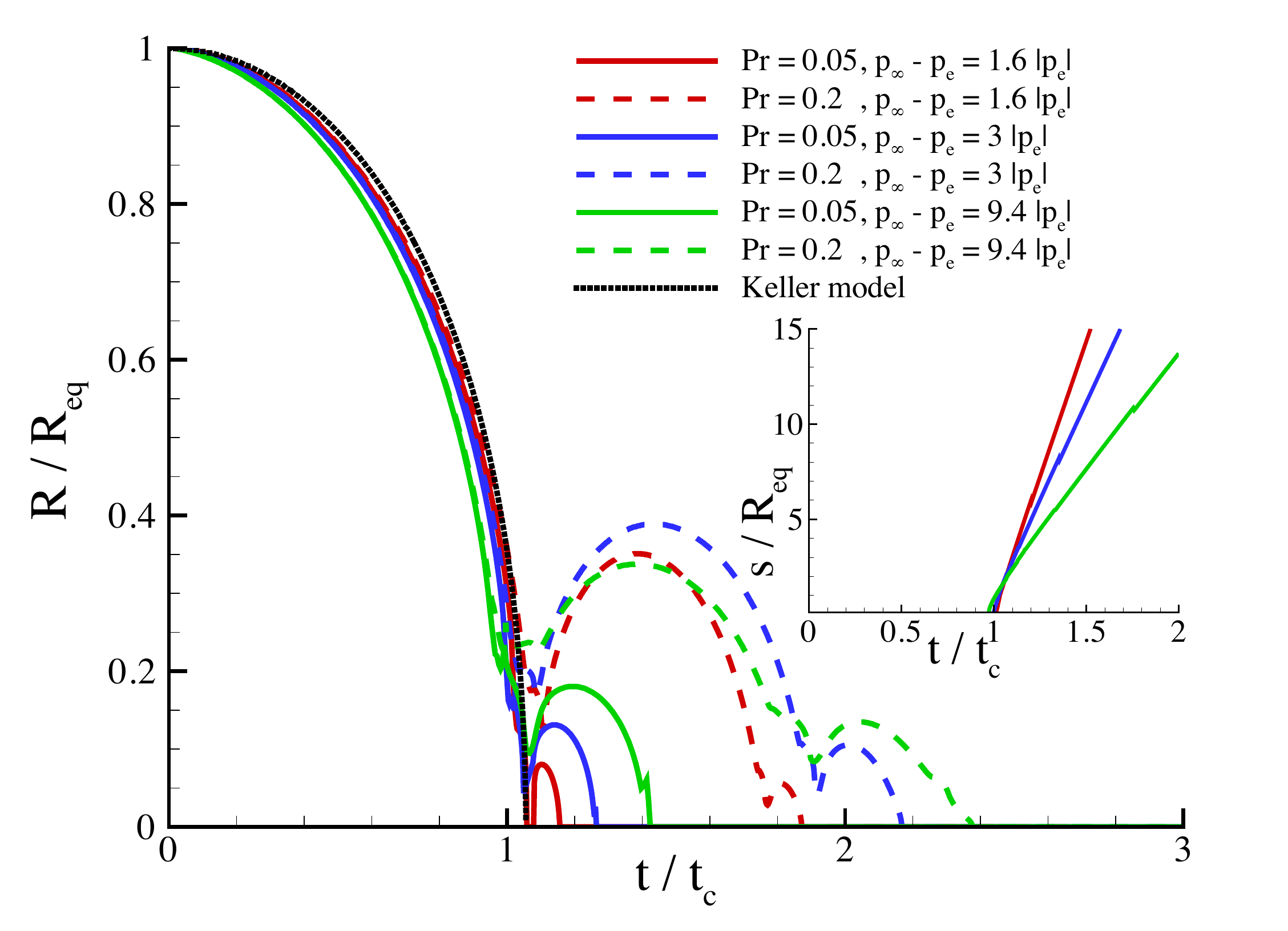}} \\
{\includegraphics[width=0.5\textwidth]{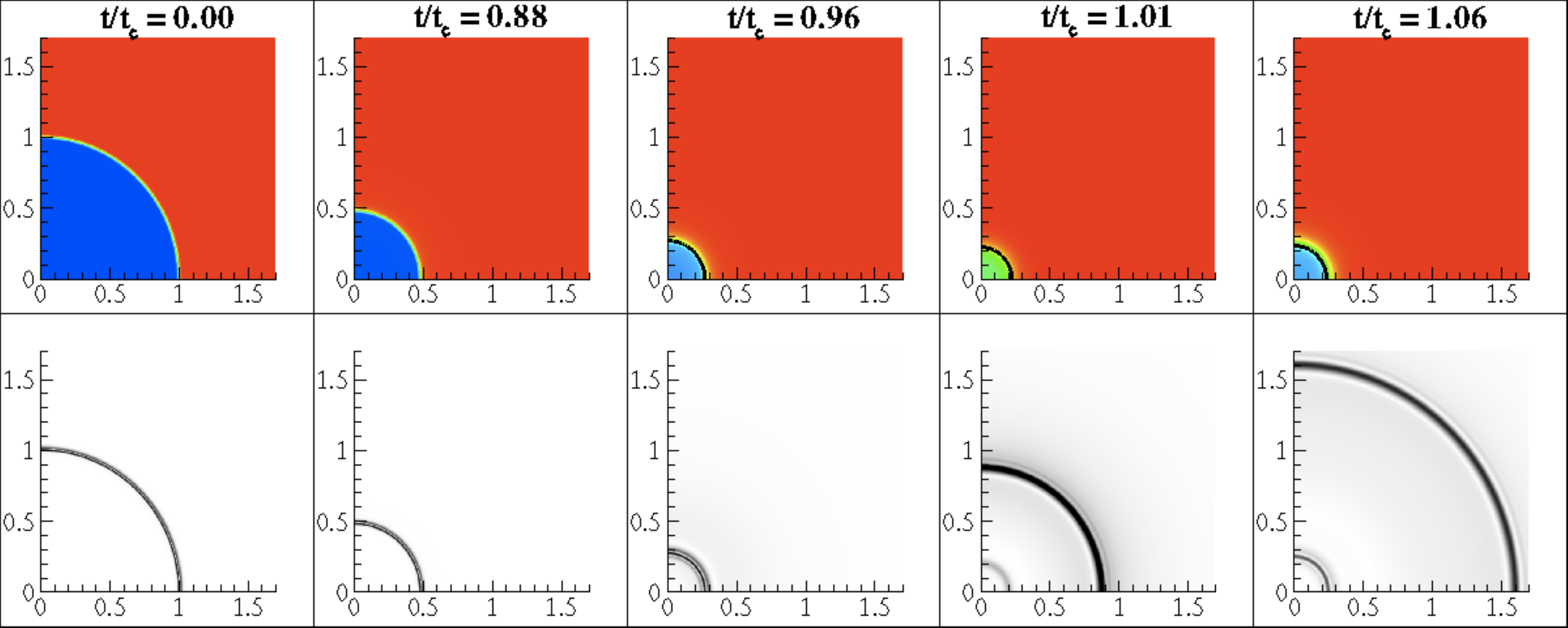}}
 \caption{Top: Time evolution of the bubble radius for various cases with the prediction of the Keller model \cite{keller} shown by the dotted line.
The collapse of the bubble is induced by an overpressure in the liquid, $\Delta p = p_\infty - p_e$.
The other control parameters are $Pr$ defined in the text, a measure of thermal conductivity, and $Re = \sqrt{p_c \rho_c} R_{eq}/\mu$ and
${\cal C} = \lambda \rho_c^2/\left(p_c R_{eq}^2 \right)$, playing the role of a Reynolds and a Weber number, see {\sl Supplemental Material, Section A} for their detailed definition. In correspondence with the first bubble rebound a shock wave is radiated in the liquid. The shock position $s(t)$ is shown in the inset. Bottom: Successive snapshots of the system configuration. Density field (upper row)  and pressure gradient intensity (lower row) in arbitrary units for $p_\infty-p_v =0.01\, p_c$ and $Pr = 0.2$.  The plots in the lower row highlight the position of bubble interface and radiated shock.
\label{f:R-t}}
 \end{figure}
see Fig.~\ref{f:R-t} where data are shown for different $\Delta p/p_e$ and $Pr$. 
The dynamics consists of a sequence of rebounds and collapses associated with shock formation. The collapse time of a macroscopic bubble is estimated as $t_c = 0.915 R_{eq}\sqrt{\rho_\infty/(p_\infty-p_v)}$, with $p_v$ the bubble equilibrium pressure, where capillary, viscous  and compressibility effects are neglected \cite{brennen2013cavitation, plesset_review}. For nano-bubbles, however, surface tension is crucial  and the numerical results suggest the  scaling
\begin{equation*}
\label{eq:tc}
t_c = 0.915 R_{eq}\sqrt{\frac{\rho_\infty}{p_\infty-p_v+2\sigma/R_{eq}}} = 0.915 R_{eq}\sqrt{\frac{\rho_\infty}{\Delta p} } \ .
\end{equation*}
Before the first collapse, the radius evolution is independent of thermal conductivity with a slight sensitivity to the overpressure. 
Although predicted by models of incondensable gas bubbles \cite{plesset_review}, rebounds are missed by simplified models which neglect the inner vapor dynamics.  
The rebounds are affected by  thermal conductivity and overpressure, Fig.~\ref{f:R-t}. 
The radius where the first  collapse phase ends, and the successive rebound starts, increases with the overpressure, suggesting the presence of an incondensable gaseous phase inside the bubble. 
The increase of the overpressure leads to faster dynamics, see the expression of $t_c$, resulting in increased pressure inside the bubble. 
An enhanced thermal conductivity, solid lines in Fig.~\ref{f:R-t}, reduces the subsequent oscillations of the bubble by diffusing thermal energy from the hotter bubble to the colder liquid, recovering the isothermal Keller model with no-rebounds in the limit  $Pr\rightarrow 0$. 

The shock position $s = s(t)$ is provided in the inset of Fig.~\ref{f:R-t}. The seemingly different velocity is an artifact of the $\Delta p$-dependent time scale $t_c$. In fact  $w = \dot s$ is fairly constant. 
Away from the bubble, the shock propagates in the still liquid that, after the expansion wave, relaxed back to  $p_\infty$, $\theta_\infty = \theta_e$.  The shock speed $w$ is determined by the state ahead of the shock ($p_\infty, \rho_\infty, u_\infty = 0$) and by an additional parameter, the density $\rho_b$ behind the shock, say, see \cite{courant1976supersonic} and \cite{ruggeri} for details concerning a van der Waals fluid.
The small compressibility of the liquid $(\rho_b-\rho_\infty)/\rho_\infty \ll 1$, allows  the linearization $w = c_\infty + \alpha_\infty (\rho_b-\rho_\infty)/\rho_\infty$, where  $c_\infty$ is the  unperturbed sound speed in the liquid and $\alpha_\infty(p_\infty, \rho_\infty)= \rho_\infty dw/d\rho \vert_\infty $. It turns out that $w \simeq c_\infty$ for the cases explicitly reported here (more precisely, $(w-c_\infty)/c_\infty \le 3 \% $).
 \begin{figure}[t!]
{\includegraphics[width=0.5\textwidth]{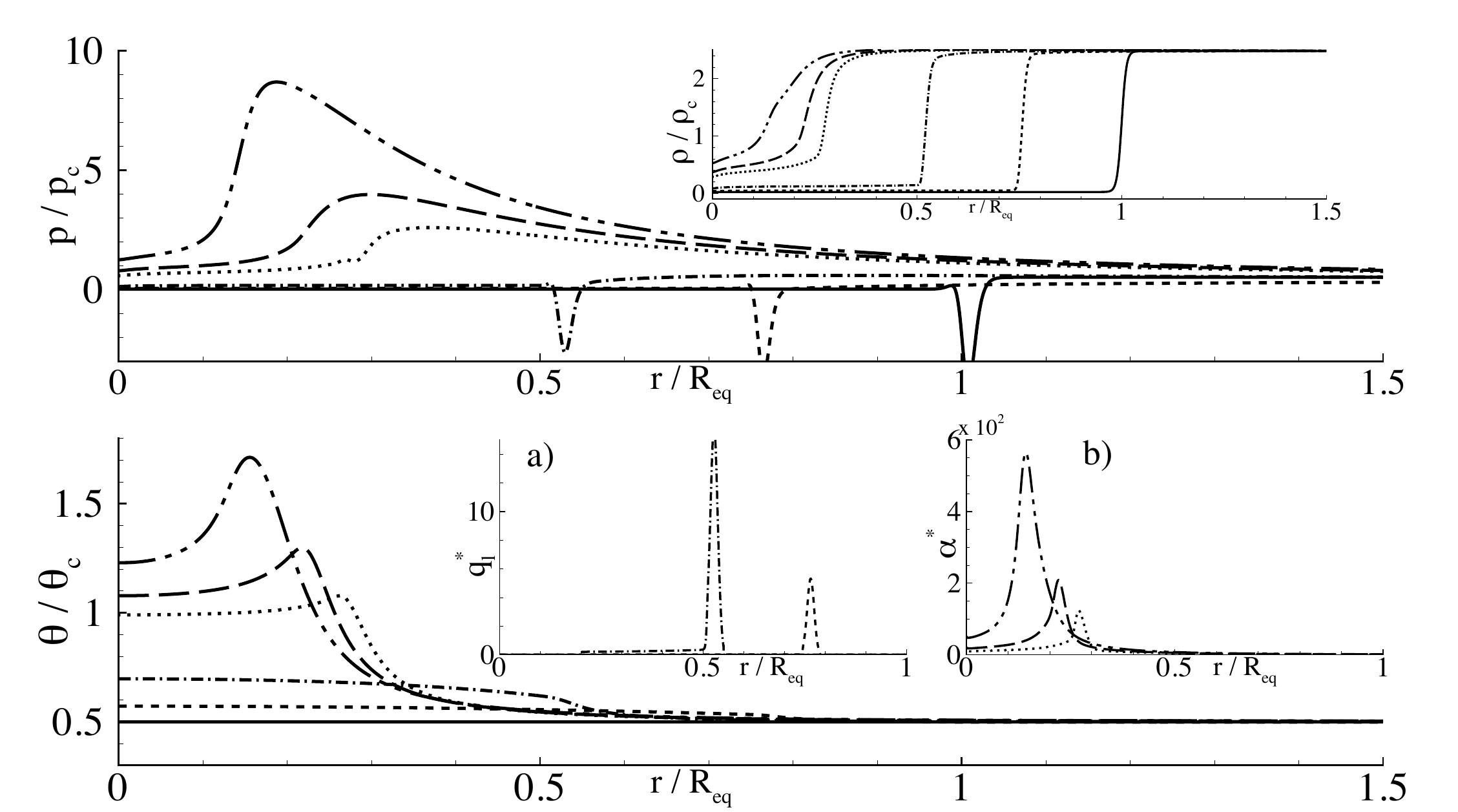}}%
 \caption{Radial profiles before the first bubble rebound. The different line styles correspond to successive time instants (solid: $t/t_c=0$, dashed: $t/t_c=0.64$, dash-dotted: $t/t_c=0.85$, dotted: $t/t_c=0.96$, long-dashed: $t/t_c=0.97$, dash-dot-dotted: $t/t_c=0.98$). The top panel shows the pressure with density in the inset. The bottom panel shows the temperature and the two contributions to the heat release in the insets: a) the non dimensional latent heat $q_l^* = {\cal{H}}q_l \dot{\rho}/(p_c^{3/2}\rho_c^{-1/2}R_{eq}^{-1})$; b) $\alpha^* = \alpha\dot{\rho}/(p_c^{3/2}\rho_c^{-1/2}R_{eq}^{-1})$.
 \label{f:prof_before}}
 \end{figure}
 \begin{figure}[h]
{\includegraphics[width=0.5\textwidth]{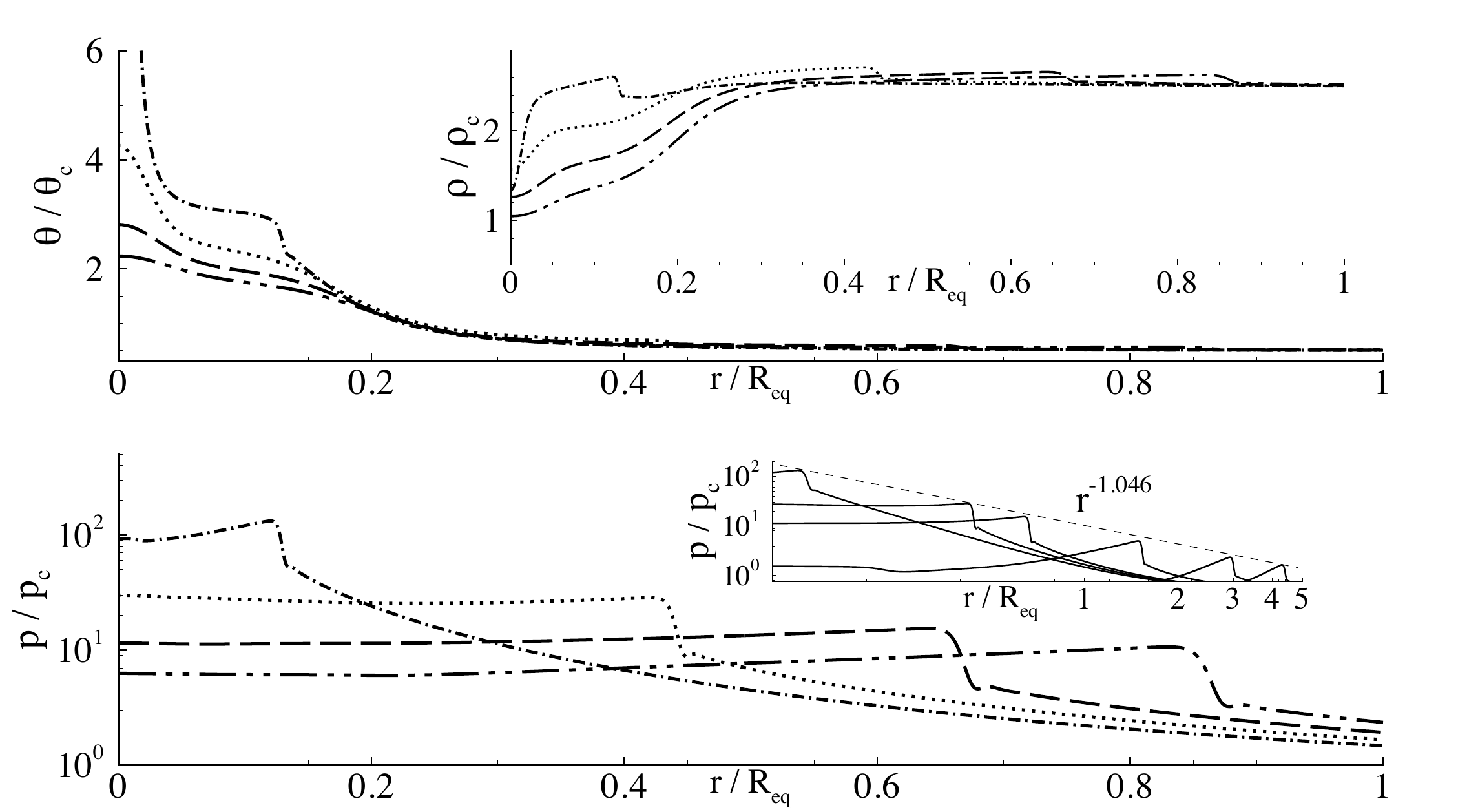}}%
 \caption{Radial profiles after the first bubble rebound. The different line styles correspond to successive time instants (dash-dotted: $t/t_c=1.005$, dotted: $t/t_c=1.01$, long-dashed: $t/t_c=1.02$, dash-dot-dotted: $t/t_c=1.03$). The top panel shows the temperature with density in the inset. The bottom panel shows the pressure. In the inset the Hickling-Plesset power law for the shock peak attenuation is compared with the simulations.
 \label{f:prof_later}}
 \end{figure}

The main plot in the top panel of Fig.~\ref{f:prof_before} shows the pressure profile through the bubble center for the initial phase of the process up to the bubble collapse. The corresponding density profiles are provided in the inset. Initially (solid, dashed and dash-dotted lines), the location of the interface is identified by an extremely sharp density drop. The bubble shrinks while the vapor density,  pressure, and temperature  (main plot in the bottom panel) increase. Successively, the shrinkage accelerates, the gaseous phase is compressed,
its temperature raises and the fluid transitions to supercritical conditions ($p/p_c > 1$, $\theta/\theta_c >1$). 
During this phase the inner vapor core is surrounded by a shell of supercritical fluid whose density increases through a strong density gradient to eventually adjoin the external liquid.
Later, the vapor disappears altogether, transformed into a low density supercritical fluid. Subsequently pressure and temperature peaks in the profile keep increasing. 
The late collapse phase is dominated by  a strong pressure wave propagating in a homogeneous supercritical fluid which focuses into a converging shock (see {\sl Supplemental Material, Section C} \cite{SM} for the comparison with a  supercritical converging shock).  Pressure and temperature extrema are reached at the first collapse time $t_c$,  when the inner low density core disappears.

In the inset of the bottom row of Fig.~\ref{f:prof_before} we analyze the thermal aspects, Eq.~(\ref{eq.temperatura}), in particular the heat release rate $-\rho \theta (\partial \eta/\partial \rho)\vert_{\theta}\, D{\rho}/Dt$. Let us consider the splitting $- \rho \theta \partial \eta/\partial  \rho\big\vert_\theta =  {\cal H } q_{\ell} + \alpha$, where ${\cal H}$ is the characteristic function of the coexistence region (${\cal H} = 1$ for states below the coexistence  curve and $0$ otherwise). The contribution due to the phase change is $q_{\ell }= \rho \theta \left[\eta_V(\theta)- \eta_L(\theta)\right]/ \left[\rho_L(\theta)- \rho_V(\theta)\right]$ and the corresponding heat release rate ${\cal H}q_{\ell} {\dot \rho}$
is displayed in inset a) of Fig.~\ref{f:prof_before} where $\alpha {\dot \rho}$ is too small to be appreciated in the plot.
At later times, inset b), the fluid in the bubble becomes supercritical, i.e. ${\cal H} q_\ell = 0$, and the term  $\alpha {\dot \rho}$  
becomes substantial  (for ideal gases $\alpha = {\bar R} \theta$, while in general $ \alpha = \theta \beta_p/(\kappa_\theta \rho)$, where thermal compressibility and thermal expansion coefficient
are  $\kappa_\theta = - (1/v) \partial v/\partial p\vert_{\theta}$ and  $\beta_p = \left(1/v\right) \partial v/\partial \theta \vert_{p}$, respectively).

After reaching  the bubble center, the shock  is reflected and propagates toward the liquid, Fig.~\ref{f:prof_later}.  The pressure peaks ahead of the shock follows the scaling law $p_s \propto 1/s$, inset of Fig.~\ref{f:prof_later}, as predicted by the compressible  Hickling-Plesset model \cite{hickling2004collapse} for incondensable gas bubbles.
After reflection, the fluid near the bubble center expands back and reduces quickly its temperature and pressure coming back to vapor with reappearance of the bubble. The expansion continues up to a maximum radius when the motion reverts and the process repeats itself. 

 \begin{figure}[t!]
{\includegraphics[width=0.5\textwidth]{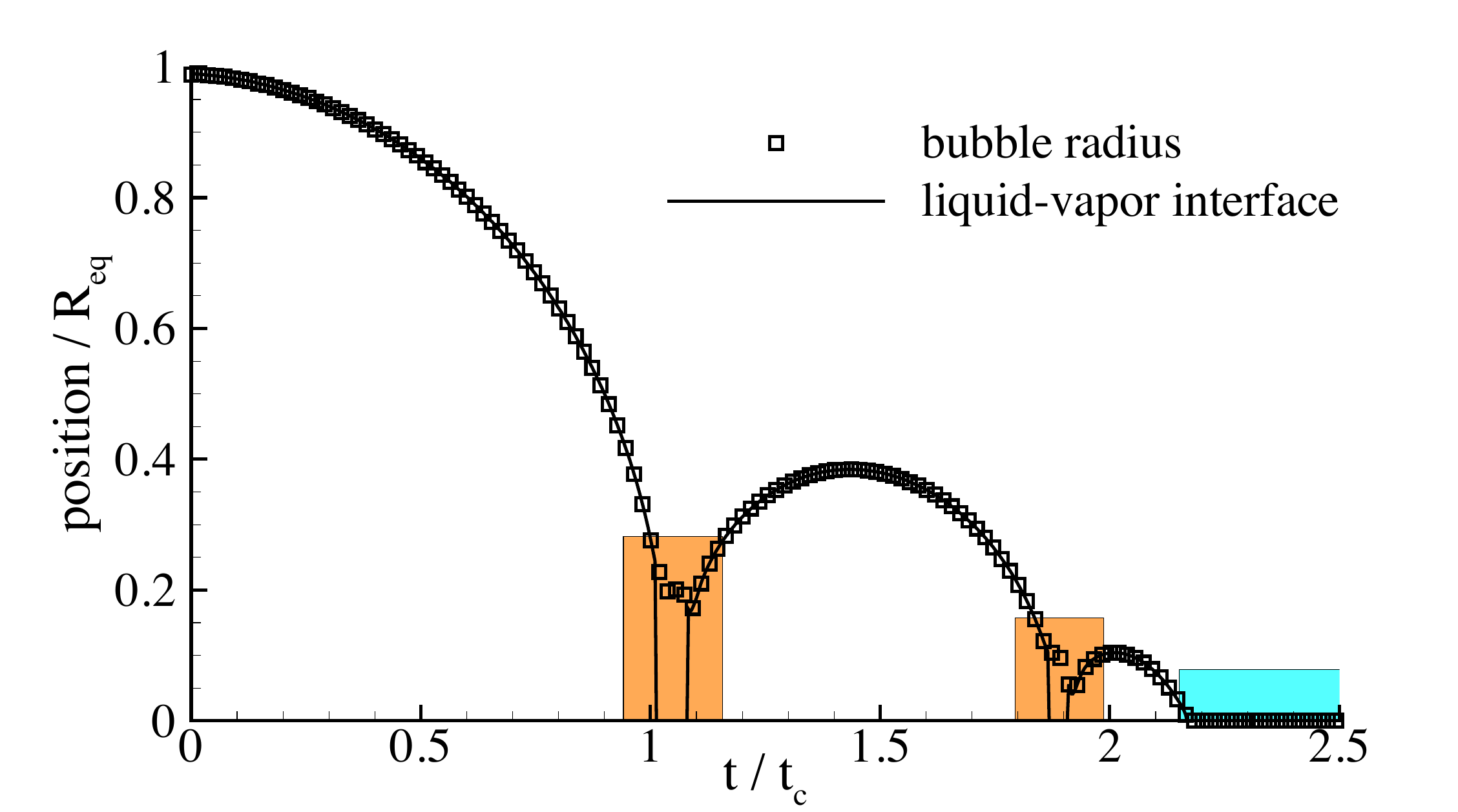}}%
 \caption{Comparison between the bubble radius and the position of the liquid-vapor interface. The bubble radius is defined irrespective of the nature of the gaseous phase inside the bubble, either the vapor or the supercritical fluid. The interface is defined as the boundary between vapor and liquid.  The bubble radius changes continuously in time, up to complete condensation and disappearance of the bubble (blue region). The interface disappears (orange regions) when the fluid inside the bubble transitions to supercritical state.
\label{f:liq-vap_interf}}
 \end{figure}
As already mentioned, during the dynamics the liquid-vapor interface may disappear and reappear several times.  In Fig.~\ref{f:liq-vap_interf} 
the interface position, defined only when vapor contiguous to the liquid exists, is compared with the bubble radius, which is well defined also when the inner fluid becomes supercritical. The time intervals corresponding to supercritical conditions are highlighted in orange. Soon after the shock emission, the inner core transitions back into vapor and the liquid-vapor interface appears again. The phenomenon repeats until a full condensation is achieved.

{\bf Discussion \& conclusions.}\label{conclusions}
\begin{table}[b]
\begin{tabular}{| c | c | c | c |}
 \hline 
  ${(p_\infty-p_e)}/{\vert p_e\vert}$ & ${p_{max}}/{p_c}$ & ${\theta_{max}}/{\theta_c}$ & ${(p_{_{r=R_{eq}}}-p_e)}/{\vert p_e\vert}$ \\
  \hline
  1.43 & 1764 & 8.58 & 31.7 \\
  0.95 & 339 & 4.53 & 22 \\
  0.63 & 190 & 3.53 & 15.9 \\
  0.073 & 52.8 & 2.01 & 6.6 \\
  \hline
\end{tabular}
\caption{
Pressure and temperature peaks in the collapsing bubble as a function of the initial overpressure. The intensity of the pressure wave at one radius from the bubble center is shown in the last column.
}
 \label{Table}
\end{table}
Phase change and transition to supercritical conditions  play a  crucial role in the collapse of a vapor nanobubble.  Indeed, independently of the intensity of the initial overpressure, a strong pressure and temperature increase is experienced that induces the transition to incondensable gaseous state. As a consequence,  a vapor bubble substantially resembles an incondensable gas  bubble,  making the boundary between the two kinds less sharply defined than usually assumed.  The pressure and temperature peaks increase with $\left(p_\infty - p_e\right)/p_e$ as the strength of the emitted shock wave does,
see Table~\ref{Table}.  At fixed thermal conductivity, a limiting overpressure exists below which the bubble condenses altogether. Above the critical overpressure oscillations set in, with the bubble periodically reforming and emitting a shock upon collapse. From the above considerations it should be expected that  a collapsing bubble could trigger a synchronized collapse of its neighbors. Indeed, the pressure at the distance $r = R_{eq}$ from the bubble center is substantially larger than the initial overpressure, Table~\ref{Table}. Accounting for the $1/r$ decay of the pressure peak, the pressure of the wave exceeds the initial overpressure in a region extending for, typically, $20 R_{eq}$.

The present model for vapor-bubbles collapse can be extended under several respects:
a) Incondensable gas dissolved in the liquid can be taken into account by extending the basic free energy functional (\ref{eq.Ffunctional}) to include the composition of the mixture;
b) More realistic transport coefficients can be assumed, e.g. dependence of viscosity and thermal conductivity on thermodynamic conditions can be included;
c) A more general EoS can be adopted  to take into account dissociation and ionization effects which are expected to quantitatively modify the dynamics of the collapse, reducing peak temperature and pressure, while maintaining the overall phenomenology basically unchanged.

{\bf Acknowledgments.} The research leading to these results has received funding from the  European Research Council under the European Union's Seventh Framework Programme (FP7/2007-2013) / ERC Grant agreement n. [339446].

\bibliographystyle{apsrev4-1}
%
\end{document}